\documentclass[%
 aip,
 apl,%
 amsmath,amssymb,
 reprint,%
]{revtex4-1}

\usepackage{epsfig}
\usepackage{dcolumn}
\usepackage{bm}
\usepackage{color}
\usepackage{xcolor}

\begin{document}


\title[Anisotropy-driven thermal conductivity switching and thermal
       hysteresis in a ferroelectric]
      {Anisotropy-driven thermal conductivity switching and thermal
       hysteresis in a ferroelectric}

\author{Juan Antonio Seijas-Bellido}
\affiliation{Institut de Ci\`encia de Materials de Barcelona (ICMAB--CSIC)
             Campus de Bellaterra, 08193 Bellaterra, Barcelona, Spain}

\author{Jorge \'I\~{n}iguez}
\affiliation{Materials Research and Technology Department, Luxembourg
             Institute of Science and Technology (LIST),
             Avenue des Hauts-Fourneaux 5, L-4362 Esch/Alzette, Luxembourg}
\affiliation{Physics and Materials Science Research Unit, University of Luxembourg,
              41 Rue du Brill, L-4422 Belvaux, Luxembourg}

\author{Riccardo Rurali}
\email{rrurali@icmab.es}
\affiliation{Institut de Ci\`encia de Materials de Barcelona (ICMAB--CSIC)
             Campus de Bellaterra, 08193 Bellaterra, Barcelona, Spain}

\date{\today}

\begin{abstract}
We present a theoretical proposal for the design of a thermal switch based
on the anisotropy of the thermal conductivity of PbTiO$_3$ and of the possibility
to rotate the ferroelectric polarization with an external electric field.
Our calculations are based on an iterative solution of the phonon
Boltzmann Transport Equation and rely on interatomic force constants
computed within an efficient second-principles density functional
theory scheme. We also characterize the hysteresis cycle of the
thermal conductivity in presence of an applied electric field
and show that the response time would be
limited by speed of the ferroelectric switch itself and thus can
operate in the high-frequency regime.
\end{abstract}


\maketitle 


A longstanding goal~\footnote{footnotes working fine} of {\it phononics}~\cite{LiRMP12,VolzEPJB16,
ZardoCOGSC19} --the discipline that studies 
phonon manipulation-- is the implementation of a phonon-based 
Boolean algebra, where operations rely on a high and a low 
conductance state, which are used to encode the logical 
values of $1$ and $0$. A fundamental requisite to perform logic 
operations is having access to two conductance states that are as 
different as possible and being able to commute back and forth 
between them.

Many materials are anisotropic and thus provide naturally access 
to such distinct conductance states, but practical implementations are 
nevertheless hindered. On the one hand, the anisotropy is often 
small, like in wurtzite crystals~\cite{TogoPRB15,Raya-MorenoAPL17,
Raya-MorenoPRM19} where $\kappa_{xx}/\kappa_{yy}$ may range from 
0.96 (GaP) to 1.18 (ZnSe) at 300~K.
In other materials, however, it can be much larger and a convenient 
limiting case is graphite: in that case, the in-plane thermal conductivity is four
orders of magnitude larger than the out-of-plane one~\cite{HarbAPL12}, 
along which phonon propagation is mediated by weak van der Waals forces. 
On the other hand, and more importantly, switching from one 
conductance state to the other is normally not possible, because 
the device design determines the element of the thermal 
conductivity tensor that is relevant for phonon transport
({\it i.e.} heat flows along a given crystallographic direction and
the sample cannot be rotated).

Ferroelectric perovskite oxides are anisotropic materials whose
anisotropy is determined by the off-center displacement of the 
cations with respect to the surrounding oxygen cages. This 
distortion of the lattice has an associated polarization that
can thus be reoriented or fully reversed with an external
electric field. This property allows designing materials with
tunable thermal properties by the continuous control of the 
distortion of the lattice~\cite{SeijasBellidoPRB18,TorresPRM19} 
or by {\it writing/erasing} domain walls that separate contiguous 
domains with differently-oriented polarizations~\cite{SeijasBellidoPRB17,RoyoPRM17}.
In this paper we discuss an even simpler effect, namely, how
an electric field can be used to rotate the polarization 
in a monodomain and, consequently, gain access to a different element of the
thermal conductivity tensor within a given device setup, thus
implementing a thermal switch. As we discuss below this approach 
does not require the design of complex multidomains and only 
relies on the anisotropy of the thermal conductivity in the monodomain state.

We study PbTiO$_3$, a paradigmatic oxide with a perovskite structure 
that is ferroelectric below 760~K. We calculate its ground state
structure, the harmonic and third-order anharmonic interatomic force constants (IFCs)
within second-principles density-functional theory (SPDFT) as implemented
in the {\sc SCALE-UP} code~\cite{WojdelJPCM13,GarciaFernandezPRB16}.
SPDFT reproduces the
vibrational and response properties of PbTiO$_3$~\cite{WojdelJPCM13}
with an accuracy comparable to most first-principles approaches
and it has a documented predictive power for the most important
structural, vibrational and response properties of ferroelectric 
perovskite oxides~\cite{ZubkoNature16,ShaferPNAS18}.
In particular, the second-principles model potential of PbTiO$_3$ 
used in this work~\cite{WojdelJPCM13} has been repeatedly shown 
to reproduce the experimental behavior of the material with good 
accuracy, except for a significant underestimate of the ferroelectric 
transition temperature (the model predicts 510~K), not relevant for 
our present purposes. Regarding the thermal conductivity, a central
quantity for the purpose of this manuscript, a meaningful comparison
with the experiments is not possible. The only experimental measurements 
reported to date were carried out in multidomain samples and thus
(i)~the measured $\kappa$ can be expected to be much lower than in a
monodomain sample~\cite{SeijasBellidoPRB17,RoyoPRM17,SerovAPL13,SledzinskaACSAMI17}
and (ii)~the anisotropy cannot be estimated. Nevertheless, the good 
agreement of our computed harmonic properties, {\it i.e.} the phonon
dispersion, with density-functional theory (DFT)~\cite{WojdelJPCM13,GarciaFernandezPRB16} and 
experiments~\cite{ZubkoNature16,ShaferPNAS18} suggest that our results are reliable.
The IFCs are then used as an input to solve iteratively the Boltzmann
Transport Equation (BTE) with the ShengBTE code~\cite{LiCPC14}
and the lattice thermal conductivity is obtained as
\begin{equation}
\kappa_{ij} = \sum_{\lambda} \kappa_{ij,\lambda} = C \sum_\lambda
f_\lambda (f_\lambda + 1 ) (h\nu_\lambda)^2 v_{i,\lambda}F_{j,\lambda}
,
\label{eq:kappa}
\end{equation}
where $i$ and $j$ are the spatial directions $x$, $y$, and $z$.
$C^{-1} = k_B T^2 \Omega N$, where $k_B$, $h$, $T$, $\Omega$ and $N$
are, respectively, Boltzmann's constant, Planck's constant, the
temperature, the volume of the 5-atom unit cell, and the number of
${\bf q}$-points. The sum runs over all phonon modes, the index
$\lambda$ including both ${\bf q}$-point and phonon band. $f_\lambda$
is the equilibrium Bose-Einstein distribution function, and
$\nu_\lambda$ and $v_{i,\lambda}$ are, respectively, the frequency and
group velocity of phonon $\lambda$. The mean free displacement
$F_{j,\lambda}$ is initially taken to be equal to $\tau_\lambda
v_{j,\lambda}$, where $\tau_\lambda$ is the lifetime of mode $\lambda$
within the relaxation time approximation (RTA). Starting from this
guess, the solution is then obtained iteratively and $F_{j,\lambda}$
takes the general form $\tau_\lambda (v_{j,\lambda}+\Delta_{j,\lambda})$, 
where the correction $\Delta_\lambda$ captures the changes in the 
heat current associated to the deviations in the phonon populations
computed at the RTA level~\cite{LiPRB12,TorresPRB17}.
Scattering from isotopic disorder is also included considering the 
natural isotopic distributions of Pb, Ti, and O, through the model 
due to Tamura~\cite{TamuraPRB83}.

\begin{figure}[t]
\includegraphics[width=1.0\linewidth]{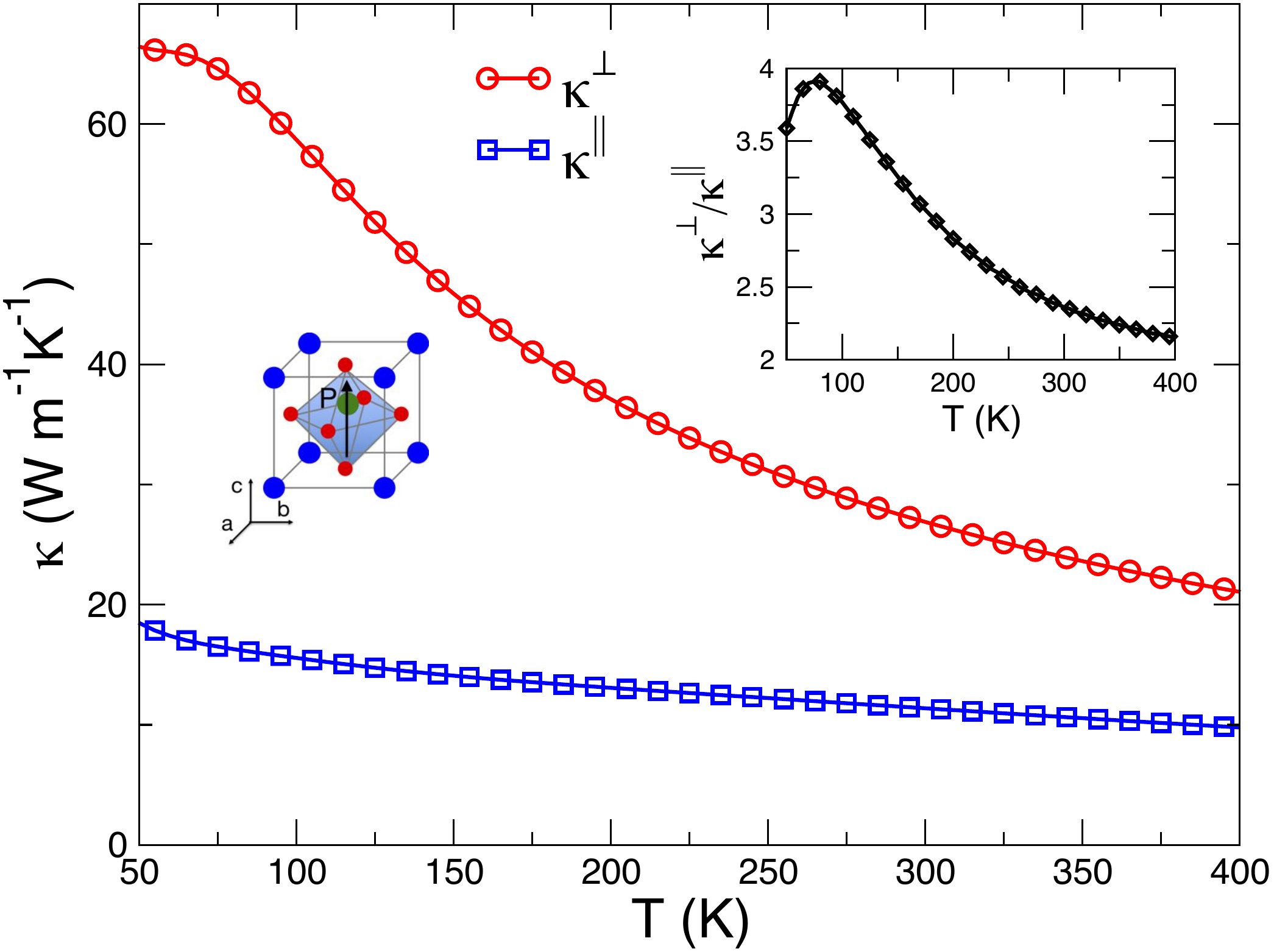}
\caption{Thermal conductivity as a function of temperature of PbTiO$_3$
         in the ferroelectric ground state along the $c$-axis and
         parallel to the polarization, $\kappa^{\parallel}$, and
         in the $ab$-plane and perpendicular to the polarization,
         $\kappa^{\perp}$. The ratio $\kappa^{\perp}/\kappa^{\parallel}$,
         which quantifies the thermal anisotropy of PbTiO$_3$, is shown
         in the inset.
        }
\label{fig:kappa}
\end{figure}

\begin{figure}[t]
\includegraphics[width=0.8\linewidth]{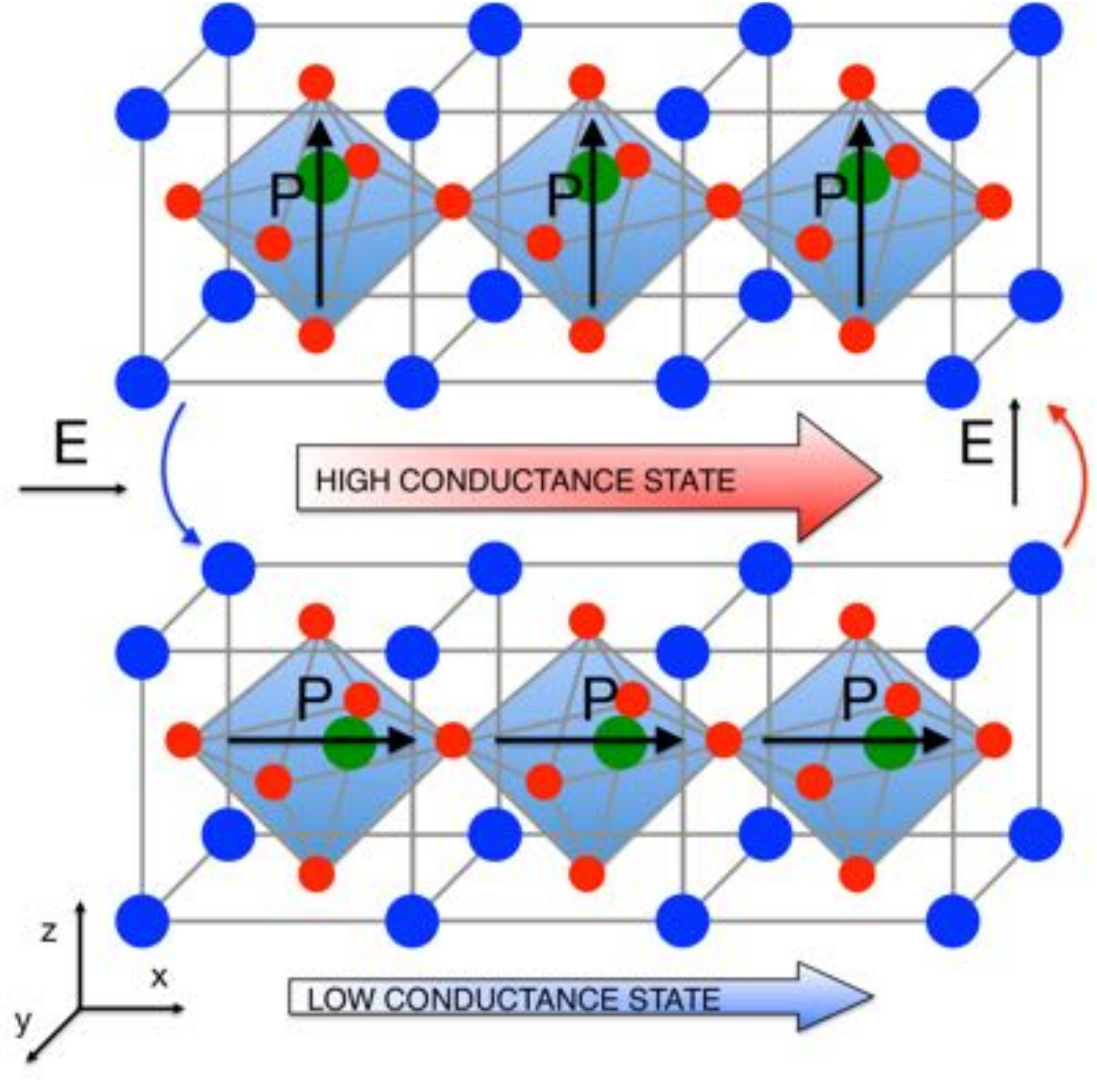}
\caption{Sketch of the thermal switch. In the high conductance state
         phonons flow along a direction perpendicular to ${\mathbf P}$
         and the thermal conductivity is $\kappa^{\perp}$. In the low
         conductance state phonon flow parallel to ${\mathbf P}$ and
         the thermal conductivity is $\kappa^{\parallel}$. An electric
         field along the $x$- or $y$-axis rotates ${\mathbf P}$ and triggers
         the commutation from high-to-low conductivity, while an
         electric field along the $z$-axis triggers the low-to-high
         conductivity transition.
        }
\label{fig:switch}
\end{figure}

The thermal conductivity tensor of PbTiO$_3$ in the ferroelectric 
ground state has two independent components and has the form

\begin{equation}
\bm{\kappa}=
  \begin{bmatrix}
    \kappa^{\perp} & 0              & 0 \\
    0              & \kappa^{\perp} & 0 \\
    0              & 0              & \kappa^{\parallel}
  \end{bmatrix}
\end{equation}

where $\kappa^{\perp}$ and $\kappa^{\parallel}$ are the thermal 
conductivities that account for heat transport when phonons flow 
perpendicular or parallel to the polarization, $\mathbf{P}$. The computed 
values of the thermal conductivity are plotted in Figure~\ref{fig:kappa}, 
where its is easy to see that $\kappa^{\perp} > \kappa^{\parallel}$ 
throughout all the temperature range investigated.
The anisotropy, $\kappa^{\perp}/\kappa^{\parallel}$, is larger at low 
temperatures (with a peak value of 
$\sim 4$ at 100~K), as shown in the inset of Figure~\ref{fig:kappa}, 
but at room temperature it is still
larger than 2. This anisotropy can be used to implement a thermal 
switch, as schematically illustrated in Figure~\ref{fig:switch}: 
phonons that propagate along the $x$-axis will experience a large
thermal conductivity, $\kappa^{\perp}$, when $\mathbf{P}$ is oriented
along $y$ or $z$, while the conductivity will be lower and equal to
$\kappa^{\parallel}$ when $\mathbf{P}$ is parallel to the heat transport
direction, $x$. Commutation between the two conductance
states is achieved with a rotation of the polarization (which is equivalent
to a rotation of the lattice/sample) by means of an electric filed that
acts as the gate control signal. 

We now take a closer look at the way the thermal conductivity, 
$\bm{\kappa}$, switches between the low and the high state. 
To this end we vary adiabatically an electric field, ${\bf E}$, perpendicular 
to the polarization to rotate the latter back and forth between 
two equivalent ground states of the PbTiO$_3$ lattice. In the limit 
of very slow variations of $\mathbf{E}$ we can assume to be always 
in equilibrium and thus we can define and calculate the thermal conductivity.
Notice that we assume this adiabatic approximation only to study the evolution
of the thermal conductivity with the electric field, {\it i.e.} with the
rotation of the polarization, and discuss its hysteretic response.
This is a not a requisite for the operation of the switch and, of course,
in an application one would like to commute between conductance states
as fast as possible. We discuss this situation below.

\begin{figure}[t]
\includegraphics[width=1.0\linewidth]{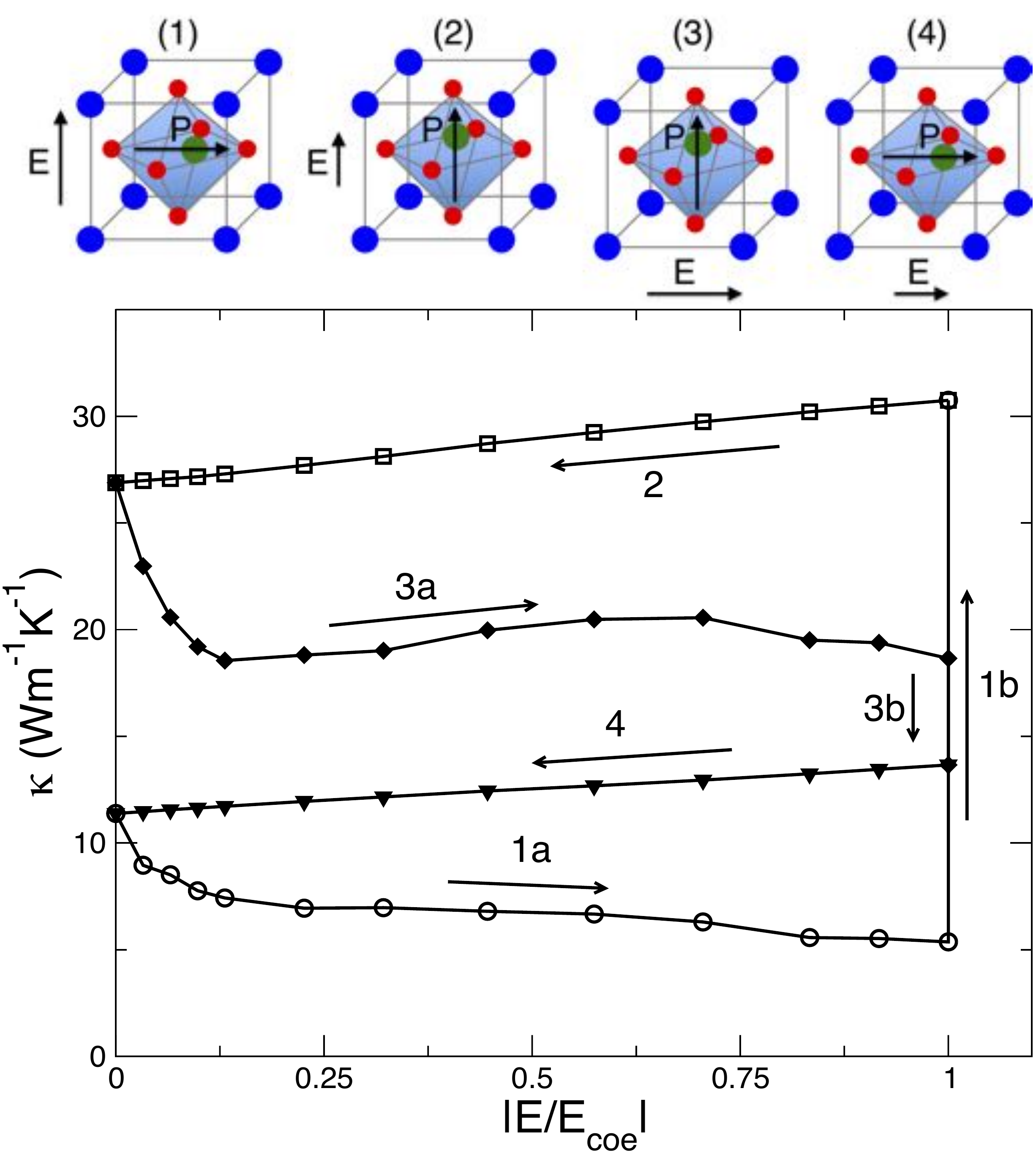}
\caption{Hysteresis cycle followed by the thermal conductivity,
         $\bm{\kappa}$, when the polarization is rotated back
         and forth at 300~K.
         The sketches in the upper part of the figure depict the
         direction of the electric field (polarization), throughout
         (at the beginning of) the corresponding branch of the
         $\bm{\kappa} (\mathbf{E})$ curve; long (short) arrows
         indicate increasing (decreasing) values of $\mathbf{E}$.
        }
\label{fig:hyst}
\end{figure}

We take as initial state the one with the polarization parallel to 
the phonon flow, that is, the low conductivity state. Then we start to 
apply a perpendicular electric field of increasing strength: $\mathbf{P}$ 
starts precessing and the conductivity is reduced (curve {\bf 1a} in 
Figure~\ref{fig:hyst}); this is a result of the field-induced lowering 
of the lattice symmetry, which increases the phase-space for three-phonon 
scattering events and thus reduce their lifetimes~\cite{SeijasBellidoPRB18}. When the coercive field, $E_{coe}$,
is reached, the polarization switches (curve {\bf 1b} in Figure~\ref{fig:hyst}) 
and becomes parallel to the electric field. If now the electric field is 
reduced until it vanishes (curve {\bf 2} in Figure~\ref{fig:hyst}) a ground state
equivalent to the starting configuration is reached. Now, however, $\mathbf{P}$
is perpendicular to the direction of phonon propagation and the system
is in the high conductivity state. To rotate back the polarization we
start applying an electric field along $x$, the direction of the heat flow: 
like before, $\bm{\kappa}$ decreases until $\mathbf{P}$ switches (curves 
{\bf 3a} and {\bf 3b} in 
Figure~\ref{fig:hyst}). Finally, upon removal of the electric field 
(curve {\bf 4} in Figure~\ref{fig:hyst}) we recover the starting 
configuration. As it can be seen $\bm{\kappa}$ follows an hysteresis 
cycle and it can take different values at a given 
electric field, depending on how the field changed in the past.
This is a direct consequence of the fact that given a value of 
$\mathbf{E}$ one cannot univocally know $\mathbf{P}$, which ultimately determines
the thermal conductivity along a given direction. 

\begin{figure}[t]
\includegraphics[width=1.0\linewidth]{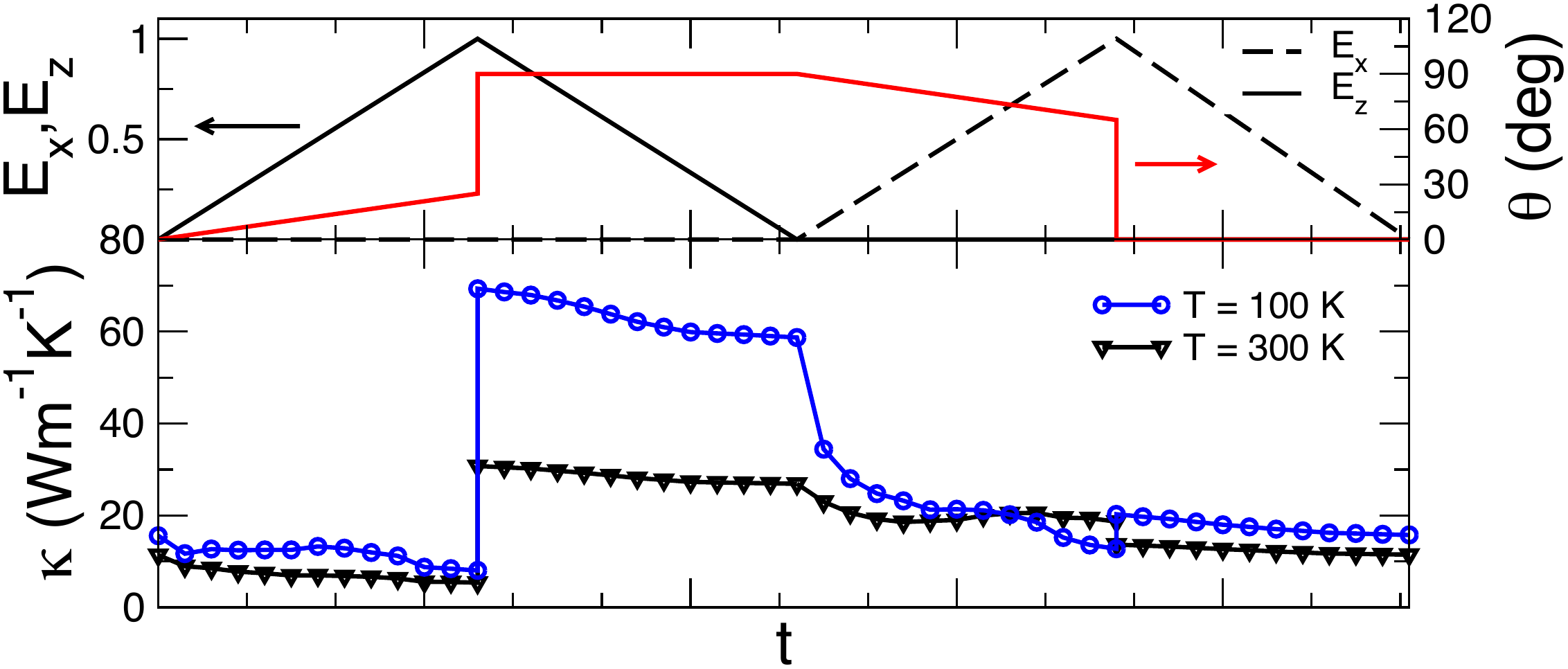}
\caption{Thermal conductivity as a function of time, $\bm{\kappa}(t)$,
         at $T=100$ and 300~K in response to the variation of the
         electric field, whose component $E_x$ and $E_z$ are plotted
         in the upper panel in unit of the coercive field, $E_{coe}$.
         The angle formed by the polarization vector
         $\theta$ and the direction of the heat flux is also shown.
        }
\label{fig:hyst_vs_t}
\end{figure}

The relation between $\mathbf{E}$, $\mathbf{P}$, and $\bm{\kappa}$ is more clearly 
illustrated in Figure~\ref{fig:hyst_vs_t} where one full hysteresis cycle
is displayed as a function of time, which is assumed to vary conveniently 
slow to allow us to consider the transition adiabatic and the system in instantaneous
equilibrium. Like in Figure~\ref{fig:hyst} $\mathbf{P}$ is initially taken to
lie along the $x$-axis, so that the system is in the low conductance
state and an electric field $E_z$ must be applied to rotate the polarization
and switch to the high conductance state. This plot highlights the
role of temperature in determining the difference between the two conductance
states, which is much larger at lower temperature, as already shown in the inset
of Figure~\ref{fig:kappa}. 

\begin{figure}[t]
\includegraphics[width=1.0\linewidth]{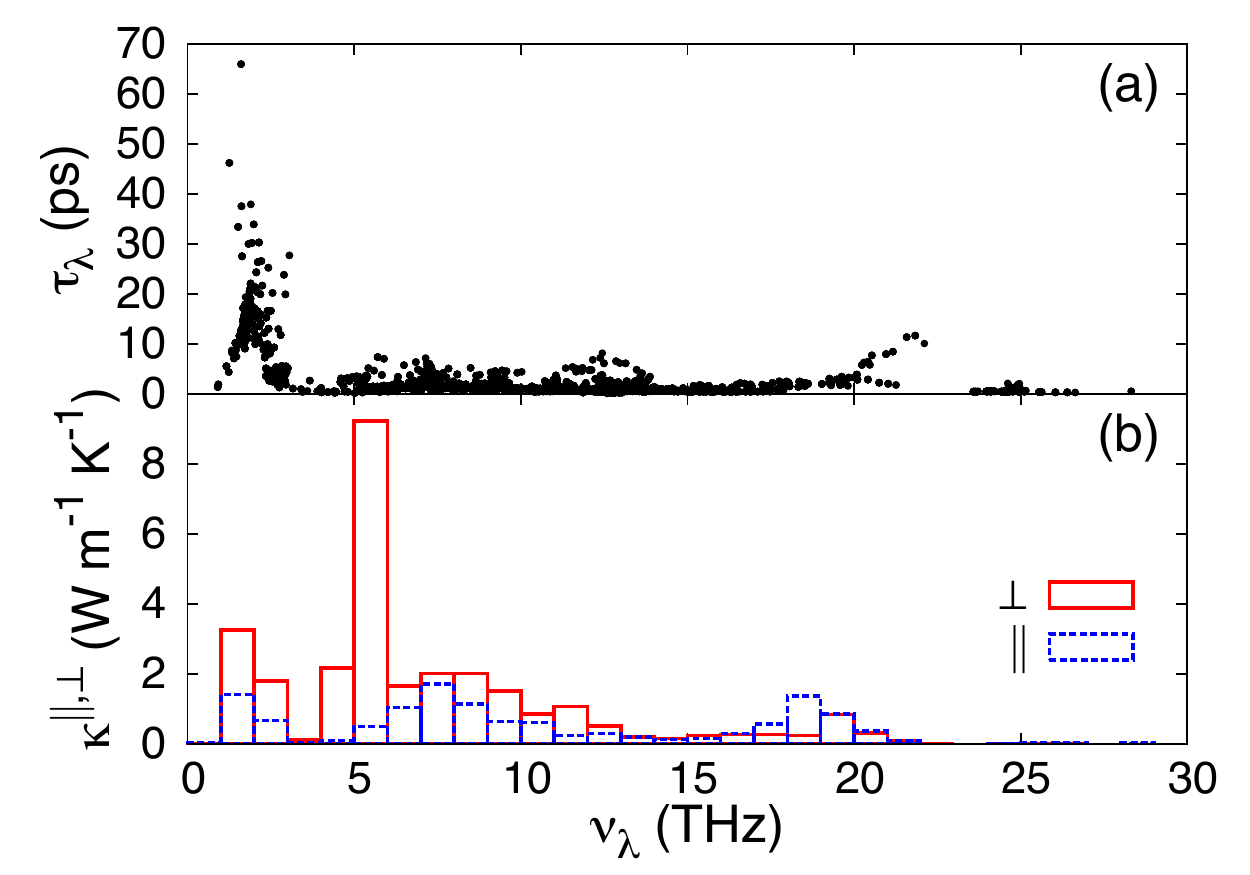}
\caption{(a)~Relaxation time, $\tau_\lambda$, as a function of frequency.
         (b)~Frequency resolved thermal conductivity,
         $\kappa^{\perp}(\nu)$ and $\kappa^{\parallel}(\nu)$.
        }
\label{fig:taus}
\end{figure}

In the discussion above the electric field was varied with slow ramps,
so that the transition could be considered adiabatic, the system
was always in equilibrium and $\bm{\kappa}$ could be computed all
along the hysteresis cycle. Of course one would like to commute 
between $1$ and $0$, i.e. high- and low-conductance, as quick as 
possible and thus would rather use short electric field pulses
to rotate the polarization. Although $\bm{\kappa}$ cannot be 
calculated in these (strongly) out-of-equilibrium conditions, we can 
infer on the overall relaxation time of the low-to-high and high-to-low 
commutation, which is an important information to estimate maximum 
operation frequency of the thermal switch. In general,
if at a given time all the modes propagate along, say, $x$ under a
certain lattice potential, and then something happens (change
of potential, removal of temperature gradient), such a propagation
will continue for a time that will depend on the phonon lifetimes.
More precisely, we introduce an effective relaxation time as
\begin{equation}
\tau^{\perp,\parallel} = \frac{1}{\kappa^{\perp,\parallel}} 
             \sum_\lambda  \tau_\lambda \kappa^{\perp,\parallel}_\lambda
\label{eq:tau}
\end{equation} 
where the relaxation time of each mode $\tau_\lambda$ is weighted with
its contribution to the thermal conductivity, $\kappa^{\perp}_\lambda/\kappa^{\perp}$ 
or $\kappa^{\parallel}_\lambda/\kappa^{\parallel}$, depending on the 
transport direction. Therefore we consider only those phonons that 
carry a significant fraction of the heat and neglect the relaxation 
times of those that do not contribution to the thermal conductivity
In this way we estimate
an upper bound of the time needed for the conductance to change from low to high and
vice versa when the polarization is rotated, giving more weight
to the modes that carry more heat.
In Figure~\ref{fig:taus}
we plot (a)~the relaxation times $\tau_\lambda$ as a function of the
frequency and (b)~the weighting factors $\kappa^{\perp,\parallel}(\nu)/
\kappa^{\perp,\parallel}$, where we grouped the mode-by-mode
contributions to $\bm{\kappa}$ in frequency intervals of 1~THz.
We have obtained $\tau^\perp=7$~ps and
$\tau^\parallel=5.8$~ps. These values are quite smaller than typical
switching times of the polarization and thus their effect of the 
switching dynamics is negligible. Therefore, taking the ultrafast 
polarization switching time of 220~ps in thin-film ferroelectrics 
reported by Li and coworkers~\cite{LiAPL04}, we can estimate a maximum
operation frequency of 4.5~GHz.

In conclusion, we present a proof-of-concept of a thermal switch
that takes advantage of the anisotropy of the thermal conductivity
in PbTiO$_3$ and, thus, of the availability of built-in low and 
high conductivity states. At variance with common anisotropic materials,
the electrically-triggered rotation of the ferroelectric polarization can be
used to switch between the two conductivity states. We also
present a detailed study of the hysteresis cycle in the response 
of the thermal conductivity, showing that its value depends on
how the electric field changed in the past.
Finally, from the computed phonon relaxation times and contributions 
to the thermal conductivity, we argue that the response of such a 
ferroelectric thermal switch will be quite fast and only limited 
by speed of the ferroelectric switch itself; hence, our proposed 
device should be able to operate in the high-frequency regime.
The proposed scheme can be extended to any material that (i)~has a 
high anisotropy of the thermal conductivity and (ii)~whose anisotropy
is directly related with a structural distortion that can be manipulated
with an electric field.

\begin{acknowledgments}
We acknowledge financial support by the Ministerio de Econom\'ia,
Industria y Competitividad (MINECO) under grants FEDER-MAT2017-90024-P
and the Severo Ochoa Centres of Excellence Program
under Grant SEV-2015-0496 and by the Generalitat de Catalunya under grant
no. 2017 SGR 1506. Work in Luxembourg was funded by the Luxembourg National Research Fund
(Grant FNR/C18/MS/12705883/REFOX/Gonzalez)
\end{acknowledgments}


%

\end{document}